\documentclass[prd,aps,floats,floatfix,eqsecnum,nofootinbib,12pt]{revtex4-1}
\usepackage{verbatim,graphicx,amssymb,amsbsy,bm,amsmath,rotating,epsfig}
\usepackage{psfrag}
\usepackage{hyperref}
\usepackage{fontenc}
\usepackage[latin1]{inputenc}
\usepackage{pstricks,pst-node,pst-text,pst-3d}
\usepackage{textcomp}
\newcommand{\be}{\begin{equation}}
\newcommand{\ee}{\end{equation}}
\newcommand{\bea}{\begin{eqnarray}}
\newcommand{\eea}{\end{eqnarray}}

\begin{document}

\title{Quantum Trans-Planckian Physics inside Black Holes \\
and its Spectrum}
\author{Norma G. Sanchez\\ 
International School of Astrophysics Daniel Chalonge - Hector de Vega,\\
CNRS, INSU, Sorbonne University, 
Paris, France}
\date{\today}
PHYSICAL REVIEW D {\bf107}, 126018 (2023)
\begin{abstract}
{\bf Abstract:} We provide a quantum unifying  picture for black holes of all masses and their main properties covering classical, semiclassical, Planckian and trans-Planckian gravity domains: Space-time, size, mass, vacuum ("zero point") energy, temperature, partition function, density of states and entropy. Novel results of this paper are: Black hole {\bf interiors} are always  {\bf quantum},  trans-Planckian and of constant curvature: This is so for {\it all} black holes, including the most macroscopic and astrophysical ones. The black hole interior trans-Planckian vacuum is similar to the earliest cosmological vacuum which classical gravity dual is the low energy gravity vaccum: today dark energy. There is {\it no} singularity boundary at $r = 0$, not at any other place: The quantum space-time is 
 {\bf totally regular}. The {\it quantum} Penrose diagram of the Schwarschild-Kruskal black hole is displayed. The complete black hole {\it instanton} (imaginary time) covers the known classical Gibbons-Hawking instanton plus a {\it new} central highly dense {\it quantum core} of Planck length radius and {\it constant curvature}.  The complete partition function, entropy, temperature, decay rate, discrete levels and density of states {\it all} include the trans-Planckian domaine. The semiclassical black hole entropy (the Bekenstein-Hawking entropy)$ (\sqrt{n})^2$ "interpolates" between the quantum point particle (QFT) entropy $(n)$ and the quantum string entropy $\sqrt{n}$, while  the quantum trans-Planckian entropy is $1/(\sqrt{n})^2$. Black hole evaporation ends as {\it a pure (non mixed)} quantum state of particles, gravitons and  radiation. 
 \\ 
Norma.Sanchez@obspm.fr,\;\;\;\qquad  https://chalonge-devega.fr/sanchez 
\end{abstract}


\maketitle

\tableofcontents

\section{Introduction and Results}

Quantum theory is more complete than classical theory and tells us what values physical observables should have.  

\medskip

Planckian and trans-Planckian domains are theoretically allowed, physically 
motivated too, such as the very early stages of the universe, as well as  the last stages of black hole evaporation, and the {\it black hole interiors}
too as we show here. 
Quantum eras in the far past universe are trans-Planckian and determine  the post-Planckian eras, e.g. inflation and the cosmological vacuum energy until today dark energy Refs \cite{NSPRD2021}, \cite{Sanchez2019},
\cite{Sanchez3}, \cite{Sanchez2}

\bigskip

Starting from quantum theory to reach the Planck scale and the trans-Planckian domain (instead of starting from classical gravity by quantizing general relativity) reveals novel results, {\it "quantum relativity"} and quantum 
space-time structure \cite{Sanchez2019}, \cite{Sanchez3}, \cite{Sanchez2}. The space-time coordinates can be promoted to quantum non-commuting operators:
comparison to the harmonic oscillator and global phase space structure
is enlighting, the  
hyperbolic quantum space-time structure generates the {\it quantum light cone} due to the relevant $[X,T]$ non-zero conmutator and  
a {\it new} quantum vacuum region beyond the Planck scale emerges.

\medskip
The space-time coordinates in the Planckian and trans-Planckian domain are no longer commuting, but they obey non-zero commutation relations: 
The concept of space-time is replaced by a quantum algebra. The classical space-time is recovered when the quantum  operators are the classical space-time continumm coordinates (c-numbers) with all 
commutators vanishing.

\medskip

{\bf In this paper} we investigate the black hole interiors, its structure and physical properties, with Planckian and trans-Planckian physics,  classical-quantum gravity duality and quantum space-time in this context. 
  
\medskip

One of the {\bf novel results} of this paper is that quantum 
 physics is a inherent constituent of  {\bf all} black 
 hole interiors, from the horizon to the center, in particular inside the most larger and astrophysical black holes. The results of this paper have  thus implications for both quantum theory and gravity and the searching of
quantum gravitational signals, for e-LISA \cite{LISA} for 
instance, after the 
success of LIGO \cite{LIGO},\cite{DESLIGO}. Not only for 
"quantum black holes", black hole evaporation and its last state huge emissions but for {\it macroscopic astrophysical} black holes. 

\medskip

 As we discuss in 
 Section II, a full complete quantum 
 theory of gravity should be a  {\bf 
 finite theory} (which is more than a 
 renormalizable theory): The 
renormalization procedure applies for the 
non-complete theories in the Wilsonian 
sense \cite{Wilson}, because they are 
valid in an own limitated  range of 
validity, 
and such known theories are not complete at the Planck scale and the trans-Planckian domain.

This framework provides in particular the gravitational entropy and temperature in the quantum trans-Planckian domaine, that is to say the extension to this domaine of the Bekenstein-Hawking entropy and Hawking temperature which are semiclassical gravity magnitudes. Interestingly, this approach applies to cosmology  too and allows a clarification of the cosmic vacuum energy or cosmological constant Refs \cite{NSPRD2021},   \cite{Sanchez2019}, \cite{Sanchez3}: The quantum $(\Lambda_Q)$ and classical 
 $(\Lambda)$ cosmological vacuum energy values dual of each other correspond precisely to the early and late universe state values respectively. \cite{NSPRD2021},  \cite{Sanchez2019},\cite{Sanchez3}.
\medskip

{\bf In this paper} we analyze the new quantum vacuum region inside 
the Planck scale hyperbolae which delimitate the quantum light cone in the Schwarschild-Kruskal  space-time.
The effect of the zero point (vacuum) quantum energy bends the space-time
and produces a constant curvature central region. We find the quantum discrete levels of the black hole space-time, and in the vacuum trans-Planckian region.
 In Section IV we describe the global quantum 
 space-time structure of the Schwarzschild-Kruskal black hole and extend the   Penrose diagram \cite{penrose} to the quantum domain. In {\bf Figure 1} we displays the new {\it quantum} Penrose diagram . 

\medskip

The quantum space-time structure is {\it discretized}
in quantum hyperbolic levels. For times and lengths larger than the Planck scale, the global space- time levels are  $(X_n , T_n) = \sqrt {2n +1}$,\; $ n = 0,1,2...$, (in Planck units), as well as  the {\it mass levels} $M_n$. 
The allowed levels cover the whole domaine from the Planck scale $(X_n, T_n) = 1$, $(n = 0)$, and the quantum 
 (low and intermediate $n$) levels until the quasi-classical and classical ones and tend asymptotically (very large $n$) to a continuum classical space-time. 
 In the trans-Planckian domain, (lengths and masses smaller than the Planck scale), in the black hole central region,  $(X_n, T_n)$ are $(1/\sqrt{2n+1})$, the most higher $n$ being the more quantum, excited and trans-Planckian ones.

\medskip

The size of the black hole is the gravitational length $ L_G $ in the 
classical/semiclassical regime, it is the quantum length $L_Q = l_P^2 / L_G$ in  the full quantum gravity regime. Similarly, for the Quantum mass $ M_Q= m_P^2/ M$, and quantum surface gravity $ K_Q = \kappa_P^2 / K_G $. Gravitational thermal features as Hawking radiation are typical of the semiclassical gravity regime. The end of evaporation is 
purely quantum and non thermal. For masses smaller than the Planck mass, the  final state {is not
\it anymore} a black hole but a  composite particle-like (or string-like) state.  Moreover, the quantum mass spectrum  for {\it all} masses we found (Section IV here), and the decay rates (Section VII) confirm this picture.

\medskip

We describe in Section V the imaginary time  manifold ({\it quantum} instanton):  The quantum trans-Planckian central core allows here to complete the classical gravity Gibbons-Hawking instanton, which is cutted at the horizon: The classical black hole instanton is regular but {\it not} complete. The black hole quantum instanton is regular and complete. In {\bf Figure 2} we depict the new quantum instanton black hole picture. 

These results allow us to describe (in Section VI) the complete Partition function covering all (classical and quantum) gravity regimes, and the trans-Planckian  entropy. We discuss the comparison between the point particle QFT entropy (without gravity), the black hole entropy and quantum strings in terms of ordered and non ordered partition numbers.

\medskip

 The discrete levels in the trans-Planckian central core of the black hole extend with decreasing $n$ from the most quantum 
highly excited levels (very large $n$) with smaller entropy $S_{Qn} = 1 / (2n + 1)$ and higher vacuum density $\Lambda_{Qn} = (2n + 1)$, until the Planck scale level $(n = 0)$. In the external black hole space-time, the  discrete levels
 extend from the Planck scale $(n = 0)$ and low $n$ to the quasi-classical and classical levels, tending  (very large $n$) to a continuum space-time. Consistently, these levels have larger gravitational 
 (Gibbons-Hawking) entropy $S_{Gn} = (2n + 1),\; n = 0, 1, 2, ...$ and lower vacuum energy $\Lambda_{Gn} =  1/(2n + 1)$. 

\medskip

 There is {\bf no singularity} at the black hole origin because: (i) The $r = 0$ mathematical singularity is 
{\bf not} physical but the result of the extrapolation of the purely
classical (non quantum) General Relativity theory, {\it out of 
its domain of physical validity}. The Planck scale and the quantum uncertainty principle in quantum gravity, precludes the extrapolation 
until the
zero length or time, which is precisely what is expected from 
quantum trans-Planckian physics: the smoothness of the 
classical gravitational 
singularities. (ii): The vacuum 
interior of the black hole is a small region of high but {\it bounded} 
trans-Planckian constant curvature and therefore
 {\bf without any singularity}.
There is {\it no}  singularity boundaries in the quantum space-time, not at $r = 0$, not at any other place. The quantum Schwarzschild - Kruskal space-time is 
 {\bf totally regular}. Moreover, the quantum hyperbolae
 $(T^2 - X^2) = \pm \sqrt {2}$ which 
replace the classical singularity $(T^2 - X^2)_{classical} (r=0) = \pm 1$, lie {\it outside} the 
allowed quantum levels $(T^2 - X^2)_{n} = (2 n + 1), \; n= 0, 1, ...$ and therefore they {\it are excluded} at the quantum level: The  singularity is {\it removed} out from  the quantum space-time. 

\medskip 
 
{\bf This paper is organized as follows:} 
In Section II we discuss why a quantum theory of graviy must be {\it finite}.
In section III we describe the classical, semiclassical and quantum Planckian and trans-Planckian black hole regions and regimes, their properties and the gravitational entropy in these three regimes. In
Section IV  we describe the quantum global Schwarschild-Kruskal space-time structure, its {\it quantum Penrose diagram},
 and the new results obtained with it. Section V deals with the black hole mass spectrum in the whole mass range, from astrophysical black holes to masses smaller than the Planck mass, passing through the Planck mass (the {\it crossing scale}). 
Sections V and VI describe the new imaginary time black hole instanton including the trans-Planckian region, the Partition function, and 
 the trans-Planckian entropy. 
 In Section VII we discuss the implications of these results for the early and last phases of black hole evaporation and the  quantum pure (non-mixed) decay rate. In Section VIII we describe the black hole interior and its quantum (trans-Planckian) de Sitter vacuum.
 Sections IX and X summarize remarks and conclusions.

\section{A Quantum Theory of Gravity must be Finite}

The construction of a complete consistent  quantum theory of gravitation continues being the greatest challenge in physics today. This is a problem of fundamental relevance 
for the quantum unificacion of all interactions and particle physics, theoretical physics and cosmology, the physical origin of the universe and its most early phases,  as well as the black hole interiors, quantum origin and end of black holes, multiverse possibilities, and several other physical implications of these problems.  

In addition, there is the possibility of "low energy" $(E << M_{Planck})$ physical effects that could be experimentally tested. One of them is the today dark energy  \cite{Riess}, \cite{Perlmutter}, \cite{Schmidt}, \cite{DES}, described as the low energy (classical, dilute, large scale) cosmological vacuum, remannent today of the high energy (quantum trans-Planckian, highly dense, small scale) cosmological vaccum at the origins Refs \cite {NSPRD2021}, \cite{Sanchez3}. 

A problem mostly discussed in connection with gravity quantization is the one of the {\it renormalizability} of the Einstein theory (or its various generalizations) when quantized as a local {\it quantum field theory} (QFT). A complete quantum theory at the Planckian and trans-Planckian domain must have the today's General Relativity, Quantum Mechanics and Quantum Field Theory as limiting cases. Physical effects combining gravitation and quantum mechanics are relevant  at energies of the order of $M_{Planck} = \sqrt{\hbar/G}  = 1.22 \; 10^{16} \; TeV$ and beyond, namely the trans-Planckian domaine:
$$ 
E_{Planck} \; \leq \;E \;< \;\infty, \;\quad 0 \;< \;L \;\leq l_{Planck}\; = \;10^{-33} \; cm.
$$ Such energies were available in the Universe at times $ 0 < t \leq t_{Planck}  = 5.4 \; 10^{-44}$ sec. 

Nevertheless, "low energy" ($E << M_{Planck}$) 
physical effects could be experimentally tested, like the today cosmological vacuum, Refs \cite{Riess}, \cite{Perlmutter}, \cite{Schmidt}, \cite{DES}.  In addition,  one may speculate about effects analogous to the presence of magnetic monopoles in some Grand Unified Theories, (monopoles can be detected by low energy experiments in spite of their large mass).

\medskip

A theory valid at the Planck scale and beyond, that is in the trans-Planckian domain \\ $E > E_{Planck} $, $L < l_{Planck}$,  necessarily involving quantum gravitation, will also be valid at any lower energy scale. One may ignore higher energy phenomena in a low energy theory, but {\it the opposite is not true}. In other words, a theory of quantum gravity will be a {\it"theory of everything"}. This conclusion is totally independent of the use or not of string models. It may not make physical sense to quantize {\it pure gravity}. A physically sensible quantum theory cannot contain only gravitons. For example, a theoretical prediction for the graviton-graviton scattering at energies of the order of $M_{Planck} $ {\it must include}  all particles produced in a real experiment, that is, in practice, {\it all} existing particles in Nature, since gravity couples to all matter.

\medskip

Let us discuss from a conceptual point of view the {\bf renormalizability question for gravity}.  As is clear from the preceding discussion, we have $ M_{Planck}  \leq \Lambda_0 < \infty $ for gravity. There cannot be any quantum field theory of particles beyond it. Therefore, if ultraviolet divergences appear in a quantum theory of gravitation, there is no way to interpret them as coming from a higher energy scale as it is usually done in QFT. That is to say, no physical understanding can be given to such ultraviolet infinities. The only logically consistent possibility would be to find a {\it finite} theory of quantum gravitation which is a "Theory of Everything" ($TOE$).

\medskip

 These simple arguments based on the renormalization group lead us to the conclusion that a consistent quantum theory of gravitation must be a {\bf finite theory} and {\bf must include all other interactions}. That is, it must be a TOE ("theory of everything"). In particular, it needs the understanding of the present desert between $1 \; TeV$ and $10^{16} \;TeV$. 

\medskip

There is an additional dimensional argument about the inference of a Quantum Theory of Gravitation $\rightarrow$ TOE: There are only three dimensional physical magnitudes in Nature: (length, energy and time) and correspondingly only three dimensional constants in nature: $ (c, h, G )$. All other physical constants like:
$\alpha \;= \;1/137,\; 04..., \; M_{proton}/m_{electron}, \;\;\theta_{W S} ,... etc $ are pure numbers and they must be calculable in a TOE. 

\medskip

The exhibit of $(c, G, h)$ helps in recognizing the different relevant scales and physical regimes.  Even if a hypothetical underlying "theory of everything" could only require
pure numbers (option three in Ref. \cite{Duff}), physical touch
at some level asks for the use of fundamental constants 
\cite{Okun}, \cite{Kuchar}, \cite{Gibbons}, \cite{Sanchez2003}. Here we use
three fundamental constants, (tension being $c^2/G$). It appears from our study here and in Refs \cite{NSPRD2021},  \cite{Sanchez2019}, \cite{Sanchez3}, \cite{Sanchez2} that a complete quantum theory 
of gravity is a theory of {\it pure numbers}.

\section{Classical, Semiclassical and Quantum Black Holes}

The physical classical, 
semiclassical and quantum Planckian and trans-Planckian gravity regimes are particularly 
important for several reasons, eg: the different stages of the universe evolution, the different stages of the black hole evolution (origin, evaporation and end), the different regions of the 
 global complete (Kruskal-like completion) black hole space-times.

\medskip

{\bf (i)} The classical gravity regimes are those of classical space-time with  very low energies ($E << E_{Planck}$ and large sizes $L_G >>l_{Planck}$), semiclassical gravity is that of curved space-times with QFT for matter, back reaction included, as the cosmic inflation quasi-de Sitter stage of the universe, (with typical energy scale being 
  the Grand unification scale, not larger than it), and the black hole evaporation in its early and middle stages. Quantum gravity regime includes Planckian and trans-Planckian energies, as the early universe stage  at and before the Planck time, the last black hole evaporation stages, the quantum space-time black hole regions inside the event horizon, and more generally, the quantum space-time region inside the "quantum light-cone". 

\medskip

- The classical/semiclassical gravity regime corresponds to any of the external space-time regions outside the black hole horizon until the asymptotic far regions, as well as the early (semiclassical / semiquantum) gravity phases of the black hole evaporation. 

- The quantum black hole regimes refer to the highly small quantum trans-Planckian interior of the black hole, as well as to the highly quantum gravity last phases of Black Hole evaporation.  

- For {\it any} black hole,  the classical or semiclassical gravity regimes and the quantum  (Planckian and trans-Planckian) gravity regimes  are 
{\it classical-quantum duals} of each other 
in the precise sense of the classical-quantum duality.
This means the following:

- The classical/semiclassical Black Hole $(BH)_{G}$, (that is, large black hole sizes and masses, external black hole regions),  is clearly characterized by the set of physical gravitational 
magnitudes or observables (size,  mass, classical temperature or surface gravity, entropy) $\equiv (L_G, M, T_G, S_G)$:
\begin{equation}\label{ULambda1}
(BH)_{G}\; =\; (L_G, \;M_G, \;T_G, \;S_G)
\end{equation}
- The highly dense very quantum Black Hole regime  $(BH)_Q$ is characterized by the corresponding set of quantum dual physical quantities $(L_Q, M_Q,  T_Q, S_Q)$ in the precise meaning of the classical-quantum duality:
\begin{equation} \label{UQ1}
(BH)_Q\;= \;(L_Q,\; M_Q,\;  T_Q,\; S_Q)
\end{equation}
\begin{equation} \label{Udual1}
(BH)_Q \;= \; \frac{(bh)_P^2}{(BH)_G}, \qquad \; (bh)_P \;=\; (l_P,\; m_P,\;  t_P,\; s_P)    \end{equation} 
\;
$(bh)_P$ standing for the corresponding quantities at the fundamental constant Planck scale,
the {\it crossing scale} between the two main, classical and quantum, gravity domains. 

\medskip

The black hole horizon 
 separates the interior region which is quantum and trans-Planckian from the external space-time regions which are classical and semiclassical with energies lower than the Planck energy. 
 The classical $(BH)_{G}$ and quantum $(BH)_Q$ Black Hole regimes (classical/semiclassical phases of black holes, and their quantum Planckian and trans-Planckian interior, or their very late phases of evaporation), satisfy Eqs.(\ref{ULambda1})-(\ref{Udual1}). 

The {\it total or complete} Black Hole $(BH)_{QG}$, is composed by their classical/semiclassical external regions and their quantum interior:
\begin{equation} \label{Utotal1} (BH)_{QG}\; = \; BH \left[\;(bh)_P, \;(BH)_Q, \; (BH)_{G} \right]
\end{equation} 
The subscript $G$ stands for the classical gravitation magnitudes or domain, $Q$ stands for the Quantum ones, and $P$ for their fundamental Planck scale constant values. We will see it explicitely in the following Sections: In section IV, for the black hole regions and different regimes, and for the QG black hole   properties and physical magnitudes:  surface gravity, black hole instanton, temperature,  partition function,  density of states,  entropy, decay rates.  

\medskip

The quantum black hole  $(BH)_Q$ is generated  from the classical black hole 
$(BH)_{G}$ through Eqs.(\ref{ULambda1})-(\ref{Utotal1}): {\bf classical-quantum black-hole
duality}. The {\it complete} (classical plus quantum) black-hole $(BH)_{QG}$ endowes a  {\it classical-quantum black hole CPT symmetry}. 
This includes in particular the classical, quantum, and total black hole temperatures and entropies 
and allows to characterize in a  precise way the different 
classical, semiclassical, Planckian and trans-Planckian black hole domains. 

\medskip

The black hole size is the gravitational length $L_G $
in the classical regime, it is its quantum length $L_Q = l_P^2 / L_G $ in the quantum dual regime (which includes the full quantum Planckian and trans-Planckian regime). The   {\it complete}\; size $L_{QG}$  endowes the symmetry $Q \leftarrow \rightarrow G : (L_G/l_P) \leftarrow \rightarrow (l_P/L_G)$. 
The complete (QG) (classical and quantum) variables, in particular the length $L_{QG}\;(l_P, L_G) $ cover the {\it complete} black hole  
 manifold including the quantum trans-Planckian interior and the semiclassical and classical black hole exterior. 
 {\bf(i)} {For  $m_P < M \leq \infty:\;  L_{QG} \simeq L_G, \; \; L_G > L_Q$}, which is the classical or semiclassical gravity domain.  {\bf(ii)} {For $0 \leq M < m_P: \; L_{QG} \simeq L_Q,  \; L_Q > L_G$,} which is the standard elementary particle physics domaine.
{\bf(iii)} {For $M = m_P :  L_{QG} = 1 = L_Q = L_G = l_P$}, it is the Planck scale (the {\it crossing scale}).

\medskip

Similarly, the horizon acceleration (surface gravity) $K_{G} = 
c^2 / L_G$ of the black hole in its classical gravity regime 
becomes the quantum acceleration $K_Q = k_P^2/ K_G$ in the 
quantum dual gravity regime. The classical temperature $T_{G}$, 
measure of the classical gravitational length or mass (in units 
of $\kappa_B$), becomes the quantum temperature $T_Q$ (measure 
of the quantum size or Compton length) in the quantum  regime. 
{\it Consistently}, the Gibbons-Hawking temperature  
is precisely the quantum temperature $T_Q$. 

\medskip

Similarly, the classical/semiclassical gravitational area or 
entropy $S_G$ (Bekenstein-Hawking entropy) has its quantum dual 
$S_Q = s_P^2 / S_G $ in the quantum gravity (Planckian and 
trans-Planckian) regime, $s_P = \pi \kappa_B$ being the Planck 
entropy: 
\begin{equation} \label{SrhoQLambda}
S_{G}\;= \; \frac{s_P}{4}\;\left(\frac{A_G}{a_P}\right) = s_P \;\left(\frac{M}{m_P}\right)^2
\end{equation}
\begin{equation} \label{SQrhoLambda}
S_Q \;=  \; \frac{s_P}{4} \; \left(\frac{a_P}{A_G}\right) = s_P \; \left(\frac{m_P}{M}\right)^2
\end{equation} 
The {\it total} $QG$ (classical and quantum) gravitational entropy $S_{QG}$ derives from the general expression 
$$S_{QG} \;= \; k_B \;\; \frac{A_{QG}}{\; 4 \;l_{P}^2} $$
where $A_{QG} = 4 \pi \;L_{QG}^2  = 4 \pi \; (L_Q + L_G)^2$ is the total area which expresses as = $A_{QG} = A_Q  +  A_G + 2 a_P$.          
Recall that  $L_Q = l_P^2 / L_G$   and  $a_P  =  4\pi l_P^2$. As a consequence: 
\begin{equation} \label{Stotalvalue}
S_{QG}\; = \; 2\;s_P + S_{G} + S_Q 
\;= \;2 \;s_P \;\left[\; 1 + \frac{1}{2}\; \left(\;\frac{S_G}{s_P} + \frac{s_P}{S_G}\;\right)\;\right]
\end{equation}
The {\it total}\;$(QG)$ gravitational entropy is the sum of the three components as it must be: classical (subscript $G$), quantum (subcript $Q$) and Planck value (subscript $P$) corresponding 
to the tree gravity regimes. The term $s_P$ arises from the duality between the quantum and classical black hole sizes $L_Q$ and $L_G$ across the Planck scale. It reflects the complete $QG$ covering: the Planck scale being the bordering or crossing scale common to the two (classical and quantum) $Q$ and $G$ domains, and to the two black hole regions: classical (exterior) and quantum (interior) black hole regions. 

\medskip

The gravitational entropy $S_{G}$ of large (classical) large {\it astrophysical black holes} is a very {\it huge number}, 
consistent with the fact that classical black holes contain a very huge amount of information. Moreover, to reach such a huge entropy, the black hole in its 
 late collapse state should have been in a highly energetic vacuum state of amount $S_G$.

\medskip

The gravitational (Gibbons-Hawking  
\cite{GibbHawkEntropy} and Bekenstein
\cite{bekenstein1981})  entropy covers the classical/semiclassical gravity but not the fully quantum gravity domaine. 
In this domaine the relevant appropriate size of the quantum system is the Compton or quantum length $L_Q$ and not the gravitational size.  The gravitational entropies in the two
different domains are classical-quantum gravity
duals of each other. The total gravitational
entropy is the sum of the entropies in the three main gravity regimes: classical/semiclassical gravity, Planckian and Trans-Planckian regimes. 
The complete (QG) variables entail  precisely 
those three regimes, and provide the additive 
constant too, that is  the pure Planckian  scale
term (a constant). The total or complete (QG) entropy here refers to the inclusion of the quantum gravity  entropy which is trans-Planckian and corresponds to the {\bf central  quantum interior region of the black hole}. The imaginary 
 time quantum gravitational {\it instanton} treatment and the euclidean partition function we present here (in Sections V and VI below), provide further support to this entropy. 

\medskip

The {\bf complete (classical plus quantum)} physical quantities are invariant under the classical-quantum duality: 
$G \leftrightarrow Q$. As the wave-particle duality at the basis of quantum physics, the wave-particle-gravity 
duality is reflected in all black hole regions and its associated physical quantites, temperature and entropy. The classical-quantum or wave-particle-gravity duality between the different gravity regimes can be viewed 
as a mapping between the asymptotic (in and out) states characterized by the sets $BH_Q$ and $BH_G$ and thus as a Scattering-matrix description.
Recall that wave-particle-gravity duality manisfests too in the different cosmological eras and its associated 
gravity quantities, temperature and entropy, \cite{NSPRD2021}, \cite{Sanchez2019}, \cite{Sanchez3}: Cosmological evolution goes from a very early or precursor quantum trans-Planckian 
phase to a semiclassical gravity accelerated era (de Sitter inflation), then 
to the classical gravity known
eras until the present classical de Sitter phase.   

\section{Quantum Space-Time Structure of Black Holes}

The complete $QG$ variables allow to uncover that in the complete analytic extension or global structure 
of the Kruskal space-time underlies a classical-quantum duality structure: 
The external or visible region and its mirror copy  
are the classical or semiclassical gravitational domains while the internal region 
is a quantum gravitational-trans Planckian  scale-domain.
A duality symmetry between the two external regions, and between the internal and 
external parts  shows up as a {\it classical - quantum duality} through the Planck scale. External and internal 
regions show up with respect to the Planck scale hyperbolae $X^2 - T^2 = \pm1$ 
which delimitate the different black hole regions. In fact, "interior" and "exterior" lose their meaning in this region because the classical $ X = \pm T$ dissapear at the quantum level and became $X^2 - T^2 = \pm1$, (in Planck units). 

Quantum space-time can be described as a quantum  oscillator with its quantum algebra.
 From the classical-quantum duality and quantum  oscillator $(X, P)$ variables in global phase space,  the space-time coordinates are promoted to quantum noncommuting operators. In {\it classical} phase space, the mapping between Schwarzschild $(x*, p*)$, and Kruskal $(X, P)$ coordinates is given by
\be 
X =  \exp{(\kappa x*)}\cos\; (\kappa p*), \qquad \;
P =  \exp{(\kappa x*)}\sin \;(\kappa p*)
\ee
\be
(X^2 + P^2) = \exp{(2\kappa x*)} \;  = 2 \;H_{osc}, \qquad 
(X^2 - P^2) = \exp{(2\kappa x*)} \cos\; (2 \kappa p*) 
\ee
As is known, the classical Kruskal coordinates $(X,T)$ in terms of the Schwarzschild representation $(x*, t*)$  are given by
\be 
X =  \exp{(\kappa x*)}\cosh (\kappa t*), \qquad \;
T =  \exp{(\kappa x*)}\sinh (\kappa t*)
\ee
\be
(X^2 - T^2) = \exp{(2\kappa x*)} \;  = 2 \; H, \qquad 
(X^2 + T^2) = \exp{(2\kappa x*)} \cos (2 \kappa t*)
\ee
with the Schwarzschild star coordinate $x*$:
\be \label{xstar}
\qquad \exp(\kappa x*) = \sqrt{2\kappa r - 1} \;\exp(\kappa r), \quad 2\kappa r > 1
\ee
$ t*$ being the usual Schwarszchild time, $ \kappa $ is the dimensionless (in Planck units) gravity acceleration or surface gravity.
Another similar patch but with $X$ and $T$ exchanged and $x*$ defined by
$ \exp(\kappa x*) = \sqrt{1- 2\kappa r}\;\exp(\kappa r) $, holds for $2 \kappa r < 1$.

\medskip

For $(X, T)$ being quantum coordinates, ie non-commuting
operators, and similarly for $(x*, t*)$, the transformation is given by:
\be \label{QSKeqs}
X =  \exp{(\kappa x*)}\cosh (\kappa t*), \qquad \;
T =  \exp{(\kappa x*)}\sinh (\kappa t*)
\ee
\be \label{QSKeqs2} 
(X^2 - T^2) = \exp{(2\kappa x*)} \; \cosh (\kappa [x*,t*]) 
\ee
\be  
(X^2 + T^2) = \exp{(2\kappa x*)} \; \cosh (2\kappa t*) 
\ee
\be \label{QSKeqs3} 
[X,T] = \exp{(2\kappa x*)}\; \sinh (\kappa [x*,t*]) 
\ee
where we used the usual exponential operator product:\\
$\exp({A})\exp ({B}) = \exp({B})\exp({A})\exp ({[A, B]})$.

New terms do appear due to the quantum conmutators. At the classical level: 
$$ [X, T] = 0,  \qquad [x*, t*] = 0 \quad (\mbox{{\it classically}})$$ and  
the known classical Schwarzschild-Kruskal equations are recovered. 

Eqs. (\ref{QSKeqs})-(\ref{QSKeqs3}) describe the quantum Schwarzschild-Kruskal space-time structure and its 
properties. The equation for the quantum hyperbolic "trajectories" are
\be
(X^2 - T^2) = \pm \sqrt{\exp{(4\kappa x*)} + [X,T]^2 } =  
\pm \sqrt{ (1 - 2 \kappa r)^2\exp{(4\kappa r)} + [X,T]^2 }
\ee
The characteristic lines and what classically were the light-cone generating 
horizons $X = \pm T$ (at $2 \kappa r = 1$, or $x* = - \infty$) become: 
\be
X = \pm \sqrt{ \;T^2 + [X,T]^2\;} \quad \mbox {at $2\kappa r= 1$:} \quad X \neq \pm T \; ,
\; \mbox{{\it no horizons} }
\ee
 $X \neq \pm T$ at $2\kappa r = 1$ and the null horizons are {\it erased}. 
 Similarly, in the interior regions, the classical hyperbolae $(T^2 - X^2)_{classical} = \pm 1$ 
which described the known past and future classical singularity $ r = 0, (x* = 0)$ become 
{\it at the quantum level}: 
$$
(T^2 - X^2) = \pm \sqrt{\; 1 + [X,T]^2 \;} = \pm \sqrt{2}\quad \mbox {at $r= 0$:}\quad (T^2 - X^2) \neq \pm 1 \; 
 \mbox{{\it no singularity}} 
$$ 
\be
(T^2 - X^2)_{classical} = \pm 1 \quad \mbox {at $r= 0$ \; {\it classically}}
\ee
Moreover, the quantum Kruskal light-cone variables in hyperbolic space 
\be \label{UV}
U =  \frac{1}{\sqrt{2}}\;(X - T), \qquad V = \frac{1}{\sqrt{2}}\;(X + T) 
\ee
are, upon the identification $P = iT$, the $(a, a^+)$ operators in phase space: The creation and annihilation operators $(a, a^+)$ are the {\it light-cone} type quantum coordinates of the phase space $(X, P)$: 
\be \label{aa}
a = \;\frac{1}{\sqrt{2}}\;(X + i P),
\qquad a^+ = \;\frac{1}{\sqrt{2}}\;(X - i P)
\ee
The temporal variable $T$ in the space-time configuration $(X, T)$ is like the (imaginary) momentum in phase space $(X, P)$. The identification $P = iT$ yields: 
\be 
X = \; \frac{1}{\sqrt{2}}\;(a^+ + a) , \qquad  
T = \;\frac{1}{\sqrt{2}}\;(a^+ - a) \;,  \qquad [a, a^+] = 1
\ee
wich satisfy the algebra:
$$
2H = (X^2 - T^2) = (2 a^+ a + 1), \; \qquad 
 (X^2 + T^2) = (a^2 + a^{+ 2}),
$$
\be
[2H, X]  = T, \qquad \;  [2H, T]  = X, \qquad [X, T] = 1,  
\ee
\qquad \qquad $ a^+\;a = N $ being the number operator.

The quantum space-time coordinates $(X, T)$ can therefore be considered quantum oscillator coordinates
$(X, T = iP)$, including quantum space-time fluctuations with length and mass within the Planck
scale domain and quantized levels. The quadratic form (symmetric order of operators):
$$ 2 H = UV + VU = X^2 - T^2 = (2VU + 1), 
\qquad VU = N \equiv \mbox {number operator}, $$
yields the quantum hyperbolic structure and the discrete hyperbolic space-time levels:
\be \label{Xn2Tn2}
X_n^2 - T_n^2 = (2n + 1), 
\qquad n = 0, 1, ... 
\ee
The amplitudes $(X_n, T_n)$ being
\be \label{XnTn} X_n \;= \; \sqrt{2n+1}, \;\qquad T_n\;= \; \sqrt{2n+1} 
\ee
With the identification  $T = -iP$, 
the quantum coordinates $(U, V)$ for hyperbolic space-time are precisely the ($a, a^+$) operators and as a consequence $VU$ is the Number operator. The expectation value $(2n + 1)$ has a minimal non zero value for $n = 0 $ which is the zero point energy or Planck scale vacuum. 

\begin{itemize}
\item{The future and past regions to the quantum Planck hyperbolae\\
$(T^2 - X^2)_{n = 0} = \pm 1$,  {\it all} contain totally allowed levels and behaviours. There is {\it no}  singularity boundary in the quantum space-time, not at $r = 0 = x*$, not at any other place. The quantum Schwarzschild - Kruskal space-time is {\it totally regular}.}

\item{There are {\it no} singularity boundaries at the quantum level, not at $(T^2 - X^2)(2 \kappa r=1) = \pm 1 $  nor at $(T^2 - X^2) (r=0) = \pm \sqrt {2}$ . 
The quantum space-time {\it extends} without boundary beyond the Planck hyperbolae
$(T^2 - X^2)(n=0)  = \pm 1 $ towards {\it all} levels: from the more quantum (low $n$)
levels to the classical (large $n$) ones.
The black hole {\bf interior} is {\bf quantum} and trans-Planckian. The internal region to the four quantum Planck hyperbolae $(T^2 - X^2)(n = 0)  = \pm 1 $ is 
{\it totally} quantum and within the Planck scale: this is the quantum vacuum or "zero 
point energy" region of the
 {\it quantum interior} of the black hole.}

\item{The null horizons  {\it disappeared} at the quantum level.
Due to the quantum $[X, T]$ commutator, quantum $(X, T)$ dispersions and fluctuations,
the difference between the four classical Kruskal regions (I, II, III, IV)  
{\it dissapears} in the trans-Planckian domain and become one single central region.
This provides support to the {\it quantum} identification 
 at the Planck scale of the Kruskal regions, and which translates into the 
CPT symmetry at the quantum level Refs \cite{NSPRD2021},\cite{Sanchez2019},\cite{Sanchez3},\cite{ tHooft2022}}.

\item{In terms of the local, Schwarzschild variables $(x*_{n\pm}, t*_{n\pm})$ or $(x_{n\pm}, t_{n\pm})$, being 
 $x = \exp {(\kappa x*)}$, and $t = \exp {(\kappa t*)}$, the levels are:
\be \label{xntn}
x_{n\pm} = [\; \sqrt {2 \kappa r_{n\pm} - 1}\;]\; \exp{(\kappa r_{n\pm})} 
= [\;\sqrt{ 2n +1 } \pm \sqrt{2n}\;]  \ee 
\be
t_{n\pm} =  [\;\sqrt{2n+1} \pm \sqrt{(2n+1) + 1/2}\;],
\ee
$$x_{n = 0}\;(+) \;= \;x_{n = 0}\;(-) \;= \; 1: \mbox{Planck scale},$$ 
which complete all the levels. The low $n$, intermediate, and large $n$ levels describe respectively the quantum, semiclassical and classical behaviours,  and their $(\pm)$ branches consistently reflect the classical-quantum duality properties, as shown explicitely for the similar branches of the mass spectrum in this Section below}.
\end{itemize}

The classical singularity $ r = 0 = x* $ is  \textit{quantum mechanically
smeared or erased} which is what is expected in a quantum space-time 
description. 
The diagram of the global quantum  Schwarschild-Kruskal space-time, which we name the quantum Penrose diagram, is shown in {\bf Figure 1}.

\begin{figure}
\centering
\centerline{\includegraphics [height=18 cm,width=20cm] {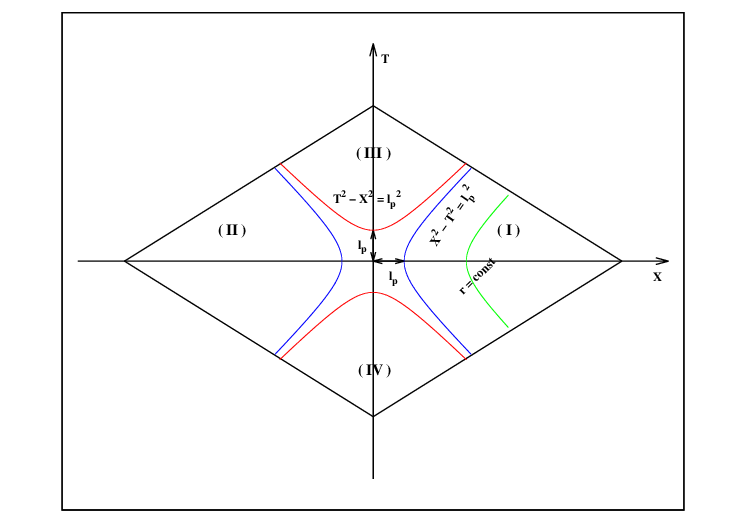}}
\setlength\abovecaptionskip{0cm}\caption{\setlength{
\baselineskip}{0.7\baselineskip}
 {\bf The quantum Penrose diagram of the Schwarschild-Kruskal
black hole}. The quantum hyperbolae $ X^2 - T^2 = \pm \; l_P^2$ replace the 
classical null horizons $X = \pm T$. The internal region to them is purely 
quantum and trans-Planckian. The difference between the four classical 
Kruskal regions (I, II, III, IV)  {\it dissapears} in the quantum domain and 
become one single central region. The exterior regions are semiclassical / 
classical asymptotically flat space-times. There is no curvature singularity 
at $r = 0$ not at any other place. The quantum space-time is totally {\bf 
regular}.  Regions extend regularly without any finite boundary nor 
curvature singularity. The central quantum region is of constant {\it 
finite} curvature. Moreover, the discrete spectrum confirms this picture: 
The quantum hyperbolae $(T^2 - X^2) = \pm \sqrt {2}$  which replace the 
classical singularity  $(T^2 - X^2)_{classical} (r = 0) = \pm 1$  lie {\it 
outside} the 
{\it allowed} quantum levels $(T^2 - X^2)_n = (2 n + 1), n = 0, 1, 2, ...$ 
and therefore, the $(r = 0)$ hyperbolae singularities are {\bf ruled out}.}
\end{figure}

\medskip

$ X_n, x_n$ in Eqs. (\ref{XnTn}), (\ref{xntn}) are given in Planck units.  In terms of the mass 
global variables  $ X = M/m_P$, or the local ones $x = m/m_P$, they translate into the mass levels:
\begin{equation}\label{Mn}
M_n = m_P\; \sqrt{(2n + 1)},\qquad \mbox{all} \;\;n = 0, 1, 2, ....
\ee
\begin{equation}\label{Mn2}
M_n \; _ {n >>1}\;= 
\; m_P \; [\;\sqrt{2\;n}\;\; + \; \frac{1}{2\sqrt{2\;n}}\; + \; O(1/n^{3/2})\;],  
\ee
\be
m_{n \pm} =  [\; M_n \;\pm \;\sqrt{ M_n^{2} - m_P^2 }\;], 
\ee
 The condition $M_n^2 \geq m_P^2$ simply corresponds to the whole spectrum $n\geq 0$:
\be
m_{n \pm} = m_P \; [\;\sqrt{ 2n +1 }\; \pm \;\sqrt{2n}\;\;]
\ee

\begin{itemize}
\item{The  quantum mass levels here holds for {\it all masses} and not only for black holes. 
Namely, the quantum mass levels are associated to the quantum space-time structure. 
Space-time can be parametrized by {\it masses} ("mass coordinates"),  
just related to length and time, as the QG variables, 
on the same footing as space and time variables.
.}  

\item{Branch (+) covers all macroscopic and astrophysical black holes as well as semiclassical 
black hole quantization $\sqrt{n}$, until masses nearby the Planck mass; and branch (-) covers quantum masses $1/\sqrt{n}$ in the Planckian and trans-Planckian domaine.}

\item {The black hole $m_P\sqrt {n}$ mass quantization is like the string mass quantization $M_n = m_s  \sqrt{n}$,  $n = 0, 1, ...$ with the Planck mass $m_P$ instead of the fundamental string mass $m_s$,  ie 
 with $G/c^2$ instead of the string constant $\alpha'$.} 
\end{itemize}

\section{Imaginary Time. The New Trans-Planckian Black Hole Instanton}

In the classical (non-quantum) Schwarschild-Kruskal space-time, taking imaginary time $ T  = i \mathcal{T} $, $ t = i \tau$, transforms the hyperbolic space-time structure into a circular structure: The characteristic lines $ X^2 + T^2 = 0 $ collapse to $ X = \pm \mathcal{T} = 0$. 
Therefore,  the  classical horizon  $X = \pm T \; (2\kappa r = 1)$ collapses to the origin, and in the classical (non-quantum) black hole {\it instanton}, the black hole interior {\it is cutted}, no horizon, and no curvature $r = 0$ singularity, does appear. 
Therefore, the {\bf classical} black hole instanton is {\it regular} but is {\it not complete}:
The interior black hole region is {\bf not} covered by the imaginary time classical (non quantum) black hole manifold.

\medskip

In the quantum Schwarzschild imaginary-time  manifold, the quantum trans-Planckian region 
corresponds to the black-hole {\it interior}, {\bf Figure 2}. Moreover, the quantum manifold covers consistently and {\it regularly} without any singularity, (not at 
$r = 0$, nor at any other place),  {\it both}: the  external and internal black hole regions. This is so in {\it both}:  The hyperbolic (real time) and  the euclidean (imaginary
time) manifolds, because of the quantum non-zero commutators $ [X,T]
$ and  $[X, \mathcal{T} ]$ respectively.

The complete  {\it quantum} black hole 
instanton includes the usual classical/semiclassical 
black hole instanton for radius larger than the Planck 
length, {\it plus a new central} highly dense {\it 
quantum  core} of Planck length radius  
and high constant and {\it finite} curvature at $r = 0$, 
corresponding to the {\it black-hole interior}, which 
is {\it absent} in the non-complete (classical) 
black-hole instanton.

In the {\it quantum} instanton
Schwarschild - Kruskal manifold, Eqs. (\ref{QSKeqs}) hold  but for $T = i \mathcal{T}, \; t* = i\tau*$ and the same star coordinate $x*$:
\be 
\exp{(\kappa x*)} = \sqrt{2 \kappa r - 1} \;\exp{(\kappa r)}, \quad 2 \kappa r > 1 
\ee
being $ \kappa = (c^2 / 2 L_G) = \kappa_P\; (m_P/ 4M ) $  the gravity acceleration or surface gravity. Another similar
patch holds for $2 \kappa r < 1$ but with $X$ and $\mathcal{T}$ exchanged, (similarly for $x*$ and $\tau*$), and with $x*$ defined by
$\exp{(\kappa x*)} = \sqrt{1 - 2 \kappa r}\;\exp{(\kappa r)}$. Therefore:
\be \label{QSKinstantoneqs}
X =  \exp{(\kappa x*)}\cos(\kappa \tau*), \qquad \;
\mathcal{T} =  \exp{(\kappa x*)}\sin (\kappa \tau*)
\ee
\be
(X^2 + \mathcal{T}^2) = \exp{(2\kappa x*)} \; \cos (\kappa  [x*,\tau*]) 
\ee
\be
(X^2 -\mathcal{T}^2) = \exp{(2\kappa x*)} \; \cos (2\kappa \tau*) 
\ee
\be
[X,\mathcal{T}] = \exp{(2\kappa x*)}\; \sin (\kappa [x*,\tau*]) 
\ee
where we used the usual exponential operator product:\\
$\exp({A})\exp ({B}) = \exp({B})\exp({A})\exp ({[A, B]})$.

The euclidean (imaginary time) {\it quantum instanton} 
clearly shows the {\it new trans-Planckian} region 
because for $ 2 \kappa r = 1$, $ (X^2 + \mathcal{T}^2)$ is {\it not} zero and have Planckian radius: The equation 
for the quantum instanton "trajectories" are
\be
(X^2 + \mathcal{T}^2) = \pm \sqrt{\exp{(4\kappa x*)} +[X,\mathcal{T}]^2 } =  
\pm \sqrt{ (1 - 2\kappa r)^2\exp{(4\kappa r)} + [X,\mathcal{T}] }
\ee
 What classically was the zero radius
 $X = \pm \mathcal{T} = 0 $ at $2\kappa r  = 1$ or $x* = - \infty$, are now: 
\be
(X^2 + \mathcal{T}^2) = [X,\mathcal{T}]^2\; \quad \mbox {at $2\kappa r = 1$:} \quad X \neq \pm \mathcal{T}  = 0\; ,\; \mbox{{\it no horizons} }
\ee
We see that 
$$X \neq \pm \mathcal{T} \neq 0 \;\;\;\;at\;\;\;\; 2\kappa r = 1.$$ The classical null horizons corresponding to the origin $X = \pm \mathcal{T} = 0$ in the euclidean signature 
 space-time (instanton) are quantum mechanically {\it replaced} by the Planck circle 
$$(X^2 + \mathcal{T}^2) =  [X,\mathcal{T}] = 1.$$
{\bf Figure 2} clearly displays this picture.
That is to say, quantum theory {\it consistently extends} the 
instanton manifold: classically the instanton is {\it "cutted"} at the 
"horizon" $ r = 1 / (2\kappa)$, while at the quantum level it {\it extends beyond it}: it contains the quantum region of Planck length radius $l_{P}$, which is neccesarily trans-Planckian and is {\it absent at the classical level}. 

That means that the 
quantum and regular imaginary time manifold,  ({\it quantum gravitational instanton}), is the usual classical/semiclassical instanton for radius larger than the Planck length {\it plus a \it central} highly dense {\it quantum core} of Planck length radius, and of high finite curvature,  which is {\it absent} classically.

\begin{figure}
\centering
\centerline{\includegraphics[height=26 cm,width=13 cm]  {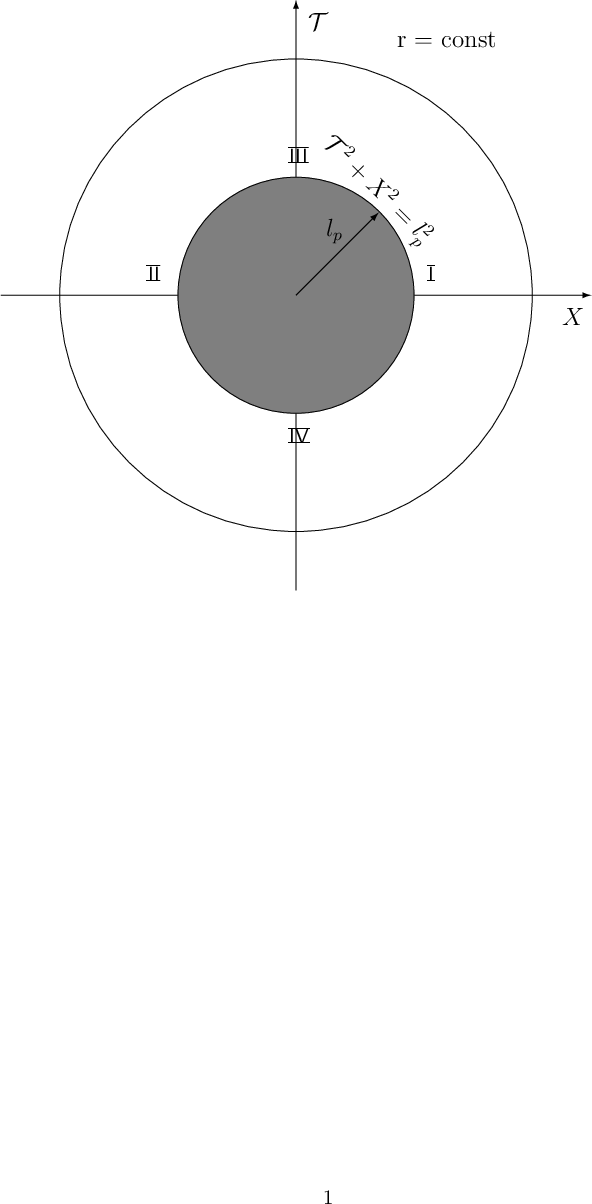}}\setlength\abovecaptionskip{-13cm} \caption{\setlength{
\baselineskip}{0.7\baselineskip}
{\bf The quantum gravitational instanton of the Schwarschild-Kruskal
black hole (imaginary time: $ T= i\mathcal{T},\; t = i\tau$)}. The classical null horizons corresponding to the origin $X = \pm \mathcal{T} = 0$ in the {\it classical} gravitational instanton of the Schwarschild-Kruskal black hole (Gibbons-Hawking instanton) are quantum mechanically {\it replaced} by the circle of Planck length radius $(X^2 + \mathcal{T}^2) =  [X,\mathcal{T}] = 1$, (in Planck units).
Quantum theory {\bf consistently extends} the 
instanton manifold: Classically, the instanton is regular but is {\it not complete} because it is {\it "cutted"} at the 
"horizon" $ r = 2M $, while at the quantum level it is {\it both}: regular and complete: The {\it quantum gravitational black hole instanton}  is the usual classical instanton for radius larger than the Planck length {\it plus a \it central} highly dense {\it quantum core} of Planck length radius, and of high finite curvature,  which is {\it absent} classically. The difference between the four Kruskal regions disappears in the euclidean manifold, they became identified. (We just indicated their places for memory of the hyperbolic manifold). The imaginary time $\tau$ in the 
 {\bf classical} instanton is {\it periodic} with period $ \beta =  2 L_G = 1 / \kappa $ : $ 1/\beta $ being the intrinsic (Hawking) temperature. In the  complete {\bf quantum} instanton, the imaginary time is periodic too but with the {\it complete} $ L_{QG} = (L_G + L_Q) $ which includes the quantum Planckian and trans-Planckian magnitudes. The complete Temperature $ T_{QG}$,  Entropy $S_{QG}$ and density of states all include the trans-Planckian domain, Section VI and Section VII. }
\end{figure}

The imaginary time $\tau$ in the 
 {\bf classical} instanton is {\it periodic} with period $ \beta =  2 L_G = 1 / \kappa_G $ : 
\be \label{beta1}
0 \;\leq \; \tau \; \leq \; \beta = 2 L_G = 1/ \kappa_G, \quad (\mbox{{\it classically}})
\ee
$ 1/\beta $ being the intrinsic manifold semiclassical temperature:  the Hawking Temperature 
\be
T_Q = t_P\; \left(\frac{l_P}{ 2 L_G} \right),
\ee

$t_P$ being the Planck temperature.
In the  complete or {\it total} {\bf quantum} instanton, the imaginary time is periodic as in Eq.(\ref{beta1}) but with the {\it complete} $ L_{QG} $ which includes the quantum Planckian and trans-Planckian magnitudes: 
\be \label{beta2}
0 \;\leq \; \tau \; \leq \; \beta = 2\; L_{QG} =  2\; (L_G + L_Q) = 1/ \kappa_{QG}, 
\ee
\be
\kappa_{QG}\; = \;\kappa_P \;({l_P}/L_{QG}),\quad \kappa_Q = \kappa_P^2/ \kappa_G,  \quad         \kappa_P\; = \; c^2/ 2\;l_P 
\ee 
\be
\kappa_{QG}\; = \;\frac{\kappa_G}{[\; 1 + (\kappa_G/ \kappa_P)^2\;]} \;= \;\frac{\kappa_Q}{[\; 1 + (\kappa_Q/ \kappa_P)^2\;]} 
\ee 

\medskip

In the classical/semiclassical gravity domaine :  $\kappa_G << \kappa_P$ it yields the usual classical surface gravity $\kappa_G$ of massive bodies with masses $ M > m_P$. For $ \kappa_Q << \kappa_P$, in the quantum domaine of masses $  M < m_P $,  (elementary particle domain), it yields the quantum $\kappa_Q = \kappa_P\; (4 M / m_P)$. The corresponding {\bf complete}  temperature being : 
\be
T_{QG} = t_P \; \kappa_{QG}/ (2\pi \kappa_P), \quad  T_Q = t_P^2/ T_G,  \quad  t_P =  m_P c^2/ (8\pi\kappa_B) 
\ee 
\be
T_{QG} = \frac{T_G}{[\; 1 + (T_G/ t_P)^2\;]} = \frac{T_Q}{[\; 1 + (T_Q/ t_P)^2\;]} 
\ee \;

For large masses, in the 
astrophysical domain: $ T_Q << 
t_P$, it yields the quantum 
Temperature $T_Q$, which is the 
Hawking temperature, as it must 
be. For small masses, $ (0 < M < m_P$ 
: $T_G << t_P$, it yields the 
usual temperature $T_G$ 
proportional to the mass, as it 
must be in the elementary particle domain. This is also manifest in the 
Partition function (Section VI below) and the
corresponding {\it complete} 
entropy. The Temperature is a measure of the length (in units of $\kappa_B$), 
$T_G = t_P \;( L_G / l_P), \; T_Q = t_P\; (L_Q/l_P)$, while the
gravitational entropy is a measure of the area. In this respect, is interesting to notice that:
\be
S_{QFT} \;=\; s_P\;  (L / l_P)^3 \; = >\;n
\ee\be
S_G \;= \; s_P \;  (L /l_P)^2 \;  = \; M ^2 \; = > \; (\sqrt{n}) ^2  
\ee\be
S_{string} \;= \;   s_P \; (L / l_P)  \; = \; M \; = > \; \sqrt{n}
\ee\be
S_{Q} \;= \; s_P \;  (l_P/ L )^2 \;  = \; M ^ {-2} \; = > \;  1/(\sqrt{n})^2 
\ee

In pure QFT without gravity the number of modes of the fields is proportional to the volume of the system (ie a box), and a short distance external cut-off is necessary, naturally placed at the Planck lenght $l_P$, because of QFT ultraviolet divergences. The string entropy $S_{string}$ is proportional to the length. 
The Black Hole gravitational entropy 
 is proportional to the area,  (whatever be $S_G$ or $S_Q$), and thus  {\it"interpolates"} between the non-gravitational entropy $ S_{QFT} $ and the string entropy $S_{string}$. $S_G$ which is the known Bekenstein-Hawking entropy exhibits its classical/ semiclassical nature, ie, $ L  >> l_P$ (equivalently,
$M_G >> m_P,\; \kappa_G << \kappa_P,\; T_Q << t_P$): 
$$S_G \;= \; s_P\; (T_G/T_Q) \;= \;(M c^2/T_Q) $$

\section{Partition Function. The Trans-Planckian Entropy}

As is known, $ D+1 $ dimensional quantum field theory with 
imaginary periodic time  $ 0 \leq \tau \leq \beta $ corresponds to a classical statistical mechanics or 
field theory with temperature  $1/ \beta$, which is used  too in the euclidean path integral of gravity, Ref. \cite {GibbHawkEntropy}, 
\be {\mathcal Z} \; = \;Tr\; \exp{(-\beta \mathcal{H})},
\ee
 with $\mathcal{H}$ being the euclidean Hamiltonian (the "evolution" generator in imaginary time, with the trace implaying periodic evolution  $ 0 \leq \tau \leq \beta $).

\bigskip

The complete (including {\it both} classical and quantum) black hole radius and 
temperature are $ L_{QG}$ and $T_{QG}$
 and are discussed in Section V.
The complete (whole range) discrete levels are discussed in Section IV and VI. Let us stress the following items about the partitions or the density of levels: 
\begin{itemize} 
\item{(i) The different types of discrete partitions depend on 
the physical nature of quantum elements considered (point particles, composite or extended quantum objects).} 
\item{ (ii) The number of partitions depends on whether one considers {\it ordered} or {\it unordered} partitions, that is to say, counting or not counting the permutations.} 
\item{ (iii) The degeneracy, the number of states corresponding to the same quantum number (whatever energy, mass, spin or other) depends on the items (i) and (ii) above.}
\item{(iv) The ensemble of all partitions considered as a Gibbs ensemble yields a thermodynamical partition.} 
\end{itemize}

Let us recall that the number $P_{o}(n)$ of {\it ordered} partitions of an integer $n$ into integers grows exponentially with $n$ :
\be
P_{o} (n) \; =\;2\;^{n - 1} = \;\frac{1}{2} \; \exp{ (n\;ln\;2)}
\ee

The number $P_{no}(n)$ of {\it non ordered} partitions of $n$,   \cite{HardyRama}   (ie without counting permutations), asymptotically for large $n$, grows exponentially with $\sqrt{n}$ :
\be
P_{no}(n)\; =   \;\frac{1}{4\; \sqrt{3}\; n}\;\exp{(\pi\sqrt{2n/3})}\;
\; [\; 1\; + O \;(\frac{log \;n}{n^{1/4}})\;]
\ee

\begin{itemize}
\item{Non-ordered partitions grow slower than the ordered ones. Naturally, the density of states and its degeneracy are smaller when the permutations are not accounted than when including the permutations.} 
\item{The non-ordered case corresponds to the density $P_{no}(n)$ of quantum composite elements (with internal structure, extended objects, strings, hadronic matter). The ordered case corresponds to point particles or quantum point oscillators. Moreover, the $\sqrt {n}$ characterizes the mass spectra of composite or extended oscillating objects, while $n$ is typical of the spectra of the punctual objects.}
\item{The existence or not of a {\it limiting temperature} in the corresponding ensembles is determined by {\it a pure number} {\it combinatorial structure}:  that is to say, by whether permutations are or not included, eg by whether partitions are ordered or unordered, eg by  whether the elements are point particles or extended objects with internal composite structure as hadrons, strings or other higher dimensional objects.} 
\end{itemize}

The total gravitational entropy $S_{QG}$ of the total or complete (classical and quantum)  black hole euclidean manifold, is the sum of the classical, quantum and Planck scale entropies:
\be
P_{QG} \; = \; e^{\;S_{QG}} 
\ee 
\be \label{SQH}
S_{QG} \; = \; 2 \;[\; s_P + \frac{1}{2}\; (\;S_G + S_Q\;)\;] , 
\ee
\begin{equation} \label{SH}
S_G = \frac{\kappa_B}{4} \;\frac{A_G}{l_P^2}, \qquad S_Q = \frac{\kappa_B}{4}\;\frac{A_Q}{l_P^2}, \qquad s_P = \frac{\kappa_B}{4} \frac{a_P}{l_P^2} = \pi \kappa_B,
\end{equation}
 
\medskip

The concept of gravitational entropy is {\it the same} for any of the gravity regimes: $ Area / 4 l_P^2$ in units of $k_B$.
For a classical object of size $L_G$, this is the classical area $A_G = 4 \pi L_G^2 $. For a quantum object of quantum size $L_Q$, this is the quantum area $A_Q = 4 \pi L_Q^2 $, (recall $ L_Q = l_P ^2 / L_G $). For the Planck length, this is the Planck area $a_P$ and $s_P = \pi \kappa_B$ is the Planck entropy :   
\begin{equation} \label{AHL}
A_G = a_P\left(\frac{L_G}{l_P}\right)^2,
\qquad
A_Q = a_P \left(\frac{l_P}{L_G}\right)^2 
 = \frac{a_P^2}{A_G},\qquad  a_P = 4 \pi\; l_P^2 
\end{equation}

\begin{equation} \label{SrhoQLambda}
S_{G} = s_P \;\frac{\rho_Q}{\rho_P} \; = \; s_P \;\left(\frac{M}{m_P}\right)^2 
\end{equation}
\begin{equation} \label{SQrhoLambda1} 
S_Q = s_P \; \frac{\rho_G}{\rho_P}\; = \;s_P \; \left(\frac{m_P}{M}\right)^2
\end{equation} 

\medskip
The {\it complete} entropy is:
\be \label{SQH2}
S_{QG} \; = \;
\; = \; 2 \; s_P \;[\; 1 \;+ \;\frac{1}{2}\; (\; A_G \; + \; A_Q\;)\;] 
\ee
and consistently, the complete partition function is
\be
\mathcal{Z}_{QG} \; = \;e\;^{S_{QG}} \; = \; {z}_P \; \mathcal{Z}_{Q} \;\mathcal{Z}_{G}  \ee

In the quantum space-time region, 
which classically corresponds to the 
interior region,  the total black hole 
entropy $S_{QG}$ is dominated by the 
Planck entropy $s_P$, the quantum 
entropy $S_Q$ being extremely low, 
minimal.  The total entropy $S_{QG}$ 
is very high in the external 
(semiclassical/classical) regions
and dominated by the Bekenstein-
Hawking entropy $S_G$ which is a 
classical or semiclassical gravity
entropy.

The discrete levels $n = 0, 1, 2, ....,$ cover all gravity regimes: from the quantum gravity (trans-Planckian and Planckian)  central black hole region to the semiclassical and classical exterior black hole regions.

In the non-trans-Planckian domain, black hole space-time levels (in Planck units)  for the distances $L_{Gn}$, vacuum energy $\Lambda_{Gn}$, and gravitational (Gibbons-Hawking) entropy $S_{Gn}$ are
\be L_{Gn} = \sqrt{(2n + 1)}, \; \; \;\Lambda_{Gn} = 1/(2n + 1),\;\; \; S_{Gn} = (2n + 1), \; \; n = 0, 1, 2, .....  \ee
 In the trans-Planckian phase $0 < r \leq l_P$, the quantum trans-Plankian levels ($Q$ denoting quantum) are: 
\be L_{Qn} = 1/\sqrt{(2n + 1)},\; \; \;\;\Lambda_{Qn} = (2n + 1),\;\; \; \;S_{Qn} = 1/(2n + 1), \; \; \; n = 0, 1, 2, ..... 
\ee 
The respective associated mass levels are:
\be
M_n = \sqrt{(2n + 1)},\qquad M_{Qn} = 
 1/\sqrt{(2n + 1)}
\ee
The density of states in the classical and quantum gravity phases are thus
\be
d_{Gn} = \exp\;{(2n + 1)} = \exp\;{(M_n )^2},\qquad d_{Qn} = \exp\;{[1/(2n + 1)]} = \exp\;{(M_{Qn})^2}
\ee
\be
d_{QG n} = \exp\;{[(2n + 1) +  1 /(2n + 1)]} = \exp\;{[M_n ^2 +\;M_{Qn}^2]}
\ee
The complete $(QG)$ density of states have both: the classical/semiclassical gravity density with the known (Bekenstein-Hawking) entropy $S_{Gn}$, and the quantum gravity density with the new trans-Planckian entropy $S_{Qn}$. As $n$ increases, the distances increase, $S_{Gn}$ increases and {\it consistently} the  black hole space-time {\it classicalizes}. In the central quantum region, $n$ decreases from the most highly central excited trans-Planckian levels, increasing $S_{Qn}$, decreasing $n$ until $n = 0$ and then increasing in the semiclassical and classical space-time. As described in Section V, the $n$-levels range over {\it all} scales from the lowest excited levels to the highest excited ones covering the twofold dual branches, classical and quantum, passing through the Planck scale, ($n = 0$), {\it the crossing scale}.

\section{Early and Last Stages of Black Hole Evaporation}

Our results here and mainly the Quantum mass spectrum in Section IV  have implications for the black hole evaporation in all its range.
$(X_n,T_n)$ are given in Planck (length and time) units. In terms of the global 
quantum gravity dimensionless length ${\cal L} = L_{QH}/l_P$ and mass ${\cal M} = M_{QH}/m_P$,  Eqs. (\ref{XnTn}) and (\ref{Mn})  translate into the discrete mass levels:
\begin{equation} \label{Ln} 
{\cal L}_n \;=\; \sqrt{(2n +1)}\; = \; {\cal M}_n ,\;\;\quad n\; =     \;0, 1, 2,....
\end{equation}

The black hole mass and radius have discrete levels $ M_{n \pm},\; L_{n \pm} $, from the most fundamental one $(n = 0)$, 
going to the semiclassical (intermediate $n$), to the classical ones (large $n$) which yield a continumm classical space-time, radius and mass, as it must be. 
This is clearly seen from the mass levels $M_{n \pm}$  Eqs. \ref{Mn}, \ref{Mn2}, (and similarly for the radius levels): 
\be
 M_{(n = 0)+} = M_{(n = 0)-} = M_{Q\;(n = 0)} = m_P,  \quad  \; n = 0 : \mbox {Planck mass}  
\ee
\be M_{n +} = m_P \; [\;2\sqrt{2\;n}\; - \;\frac{1}{2\sqrt{2 n}}\; 
+ \;O (1/n^{3/2})\;] , \;\;\;\quad  \mbox{ branch $(+)$ : masses $> m_P$}
\ee  
\be M_{n-} = \frac{m_P}{ 2 \sqrt{2\; n}}\; + \;O(1/n^{3/2}) , \quad \;\;\;
\mbox{ branch $(-)$ : masses $< m_P$ } 
 \ee
 
(i) Large $n$ levels are semiclassical tending towards a classical continuum space-time. Low $n$ are quantum, the lowest mode ($n = 0$) being the Planck scale. 
Two dual $(\pm)$ branches are present in the local variables ($\sqrt{2n+1}\pm \sqrt{2n}$) reflecting the duality of the 
large and small $n$ behaviours and covering the {\it whole} spectrum:
from the largest astrophysical masses and scales in branch $(+)$
to the quantum smallest masses and scales in branch $(-)$ passing by the Planck mass and length. 

The last stage of black hole evaporation
and its quantum decay belong to the quantum mass branch (-) with Planck scale masses and smaller 
until zero mass.
\begin{itemize}
\item{ Black hole masses belong to both branches (+) and (-):
Branch (+) covers  all {\it macroscopic} and astrophysical black holes as well as the semiclassical 
black hole quantization $\sqrt{2n +1}$  until masses nearby the Planck mass $(n = 0)$.}

\item{ The {\it microscopic   quantum} black holes, (with masses 
near the Planck mass and smaller 
 until the zero mass, ie originated as a consequence of black hole evaporation, or from Planckian and trans-Planckian primordial fluctuations), belong to the branch (-).}
\item{ The  branches (+) and (-) cover {\it all} the black hole
masses. The black hole masses in the process of black hole evaporation go 
from branches (+) to (-). Black hole ends its evaporation in branch (-) decaying as {\it a 
pure  (non mixed) quantum state}.}
\item{ Black hole evaporation is thermal in its semiclassical gravity phase 
(Hawking radiation) and it is {\it non thermal} in its last quantum stage, with a pure (non mixed) quantum decay rate.}
\item{In its last phase (mass of the order and smaller than the Planck mass $m_P$), the state 
{\bf is not anymore a black hole state}, but a pure (non mixed) quantum state, 
decaying like a quantum heavy particle. The quantum black holes decay in discrete levels, into elementary particle states, 
that is to say, pure (non mixed) quantum states with the decay rate :
\be \label{gamma}
\Gamma\; = \; \frac {g^2\; m}{num.factor}
\ee         
where $g$ is the (dimensionless) coupling constant, $m$ is the typical mass in the theory considered (the mass of the unstable particle or object) and the numerical factor often contains the relevant mass ratios in the decay process.}
\end{itemize}
The unifying formula  Eq.\ref{gamma} for quantum heavy particles \cite{HdeVNSprd} nicely encompass all the particle width decays in the standard  model (muons, Higgs, etc), as well as the {\it decay width} of topological and non topological solitons, cosmic defects and fundamental quantum strings \cite{HdeVNSprd} . 

For the last stages of quantum black holes, in terms of the discrete mass levels, the decay levels are:
$$ \Gamma_{n} \; =\; G \;\sqrt{2n + 1},$$
which is the same  $\sqrt{n}$ - dependence as for the decay $ \Gamma_{string}$ of quantum strings .

A quantum closed string in an $nth$ excited state  decays into lower excited states (including the dilaton, graviton and massless antisymmetric tensor fields) \cite{stringdecay} with a total width, given to the dominant order (one string loop) by :
$\Gamma_{string} \; = \;G \;T_s^3 / n^0 \;\approx \;G\; l_s^3 $
which can be also written as 
\be\Gamma_{string} \; = \; g^2\; m_s / n^0 
\; = \; G \; m_s /{\alpha'} n^0 \ee
$n^0$ being a numerical factor, $l_s, m_s$ and $T_s$ being the string length, mass and string temperature, ($\alpha'$ playing the role of $G/c^2$). That is, the string decay $\Gamma_s $ has the same structure as Eq.(\ref{gamma}) with $g \equiv \sqrt{G} / {\alpha'} $.

A  semiclassical black hole decays thermally, except in the last evaporation phases, as a "grey body"  at the Hawking temperature $T_Q$, the "grey body" factor being the classical black hole absorption cross section $\sigma_G $, eg the black hole area $A_G$, the mass loss rate being 
$ (dM/dt) = -\; \sigma\; L_G^2 \; T_Q^4 \;\approx \;  1 / L_G ^2 $,  
($\sigma$ being the Stefan constant). 
Therefore, the semiclassical black hole decay rate is given by
\be
\Gamma_{BH} \;  =  \; |\;\frac{d\; ln\; M}{dt} \;| \; = \;G\; T_Q^3 \;/ \;n^0\;
\approx \; G\;/ L_G^3 
\ee
 As evaporation proceeds, the black hole temperature increases until it reaches the 
 string temperature $ T_s = \hbar c/ (2\pi \kappa_B\;l_s),\;l_s = \sqrt{ \hbar\alpha'/c}$\; Refs. \cite{MRM-NGS-PRD2000}, \cite{Sanchez2003-1}, \cite{B-MRM-NGS2007},  undergoing a phase transition into a  
 quantum string or to a quantum composite state regime $T_G \rightarrow,   T_s$, $L_G \rightarrow l_s$: The black hole becomes a quantum string or quantum composite state and decays with a {\it width} 
$$\Gamma_{BH} \;\rightarrow \;G \;T_s^3 \; \approx \;
G\;/ \;l_s^3 \; \rightarrow \; \Gamma_{string}$$
 The semiclassical black hole decay rate $\Gamma_{BH}$ tends to the string decay rate $\Gamma_{s}$. Similarity between the black hole decay and  the elementary particle decay rate is achieved for quantum black holes, when the black hole enters its quantum gravity regime, eg the Planck mass at the ending phase of evaporation. 
 
 We compared here with the string case because the computations of black hole radiation in string theory Refs \cite{MRM-NGS-PRD2000},\cite{Sanchez2003-1},\cite{B-MRM-NGS2007} explicitely support this picture. And, on the other hand without any use of string theory, we find that the mass quantum discrete spectrum of black holes is similar to the mass quantum string spectrum. A similar picture holds for a quantum Planckian decaying state, a quantum composite state, (instead of a quantum decaying string state): a quantum state at the typical Planck (or trans-Planckian) temperature $ T_P$, with the Planck mass and length, $(m_P, l_P)$ instead of the string ones:
$$ \Gamma_{BH} \;\rightarrow \;G \;T_P^3  \; = \;
G\;/ \;l_P^3 \; \rightarrow \; \Gamma_P $$

 There are no quantum objects at such heavy mass as the Planck mass which would remain stable. They naturally decay  quantum mechanically in all particles, mainly gravitons and radiation. Therefore, the end of the black holes, the "remnant" states, are the last emitted particles, gravitons, and radiaton, and other elementary particles, but {\bf not stable} Planck mass objects.

\medskip

Finally, let us just point out that the whole process of black hole formation and end by evaporation can be considered in terms of a Scattering-matrix between the asymptotic states.

Black hole  (BH) formation through the gravitational collapse of a star can be  described as a S-matrix evolution ($\mathcal{S}_{\;BH}$): 
\be
\vert \; \Psi{_{BH}}\;(t)\; \; \rangle \; = \; \mathcal{S}_{\;BH}\; (t) \; \;\vert \; \Psi_{star}{(t = t_{in})} \; \rangle
\ee
It can be expressed in terms of the final star state at $t = t_{final}$, that is to say, the black hole state. 
And in general:
\be
\vert \; \Psi_{star}\;{(t)}\;  \; \rangle \; = \; \mathcal{S}_{star}\; (t) \; \vert \; \;\Psi_{star}  {(t_{in})} \; \rangle
\ee

In addition, black holes in turn  evaporate, and 
 asymptotically after enough long time, end into a gaz of particles and radiation
which eventually, under gravity and pressure evolution, forms again a star. That is to say, the initial gravitating gaz state forming a star can be the final gravitating gaz state emitted by the evaporating quantum black hole (QBH)(or at least a part of it):
\be
\vert \; \Psi_{star}\;(t_{in})\;  \; \rangle \; = \; \mathcal{S}_{star}\; (t_{in}) \; \;\vert \; \Psi_{QBH}\;
{(t_{final})} \; \rangle
\ee
Therefore,
\be
\vert \; \Psi_{star}\;{(t)}\;  \; \rangle \; = \; \mathcal{S}_{star}\; (t) \; \; \mathcal{S}_{star}\; (t_{in}) \; \vert \; \Psi_{QBH}\;
{(t_{final})} \; \rangle
\ee
It can be also expressed in terms of the initial state 
$\vert \; \Psi_{BH}\; {(t_{in})} \; \rangle$
instead of the final state $\vert \; \Psi_{QBH}\;
{(t_{final})} \; \rangle$. Therefore,
\be
\vert \; \Psi_{star}\;{(t)}\;  \; \rangle \; = \; \mathcal{S}_{star}\; (t) \; \; \mathcal{S}_{BH}\;(t)^{-1} \;\vert \; \Psi_{BH}\;(t)  \;  \; \rangle\; \;  
\ee

 This is another example of unitarity in a whole complete quantum evolution, the $S$-matrix in the whole process is unitary $S S^+ = 1 = S^+ S$. 
{\it "In Nature nothing is lost, all is transformed"} \cite{Lavoisier}.

\section{Black Hole Interior: The  Quantum Trans-Planckian de Sitter Vacuum}

 We described in Section IV the quantum space-time structure of black holes in terms of a quantum oscillator algebra with discrete hyperbolic levels $(X^2 - T^2)_n = (2n + 1), n = 0, 1, 2, ...$.  The zero point energy $ (n = 0)$ is the quantum and trans-Planckian vacuum in the central region delimitated by the four hyperbolae $ X^2 - T^2 = \pm 1$ of the Planck scale $ (n = 0)$ level. This is precisely a constant curvature de Sitter vacuum:
 The de Sitter vacuum can be described as a (inverted, ie with imaginary frequency) harmonic oscillator, the {\it oscillator constant} being \cite{NSPRD2021},\cite{Sanchez3}:
\be \label{oscdeS}
\kappa_{osc} \;= \;H^2,   \qquad  H\; = \; \sqrt{(8\pi\,G\,\Lambda)/3} \;=\; c\;/\;l_{osc}
\ee 
The {\it oscillator length} $l_{osc}$ is the Hubble radius, the Hubble constant $H = \kappa$ being the \text{surface~gravity}, as the black hole surface gravity is the inverse of (twice) the black hole radius. The description of de Sitter space-time as an 
(inverted) harmonic oscillator 
 derives from the Einstein Equations on the one hand  \cite{NSPRD2021}, \cite{dVSiebert},\cite{dVSanchez},
 and on the other hand stems more generally from the de Sitter geometrical description: 
as an hyperboloid embedded in a flat Minkowski 
space-time with one more spatial dimension : 
\be \label{dS}
- T^2 + X^2 + X_i^2 + Z^2 = L_{QG}^2  
\ee
\be
L_{QG} = (L_Q + L_G) = l_P\;(H/h_P + h_P/H), \qquad h_P = c/l_P
\ee
In the case of  {\bf Anti-de Sitter}, the description is the same but with $- T^2 + X^2 + X_i^2 + Z^2 = - L_{QG}^2$, and therefore  Anti- de Sitter background is
 associated to a real frequency {\bf (non inverted)} harmonic oscillator. Also, the propagation of fields and linearized perturbations in the de Sitter vacuum all satisfy equations which are like the {\it inverted} oscillator equations, \cite{GuthPi},
\cite{Albrecht},\cite{SanchezNPB1987},  or normal oscillators in Anti de Sitter.

Here in the black hole case, the physical magnitudes as the oscillator constant $H^2$ and typical length $ (c/ l_{osc})$ are related to the black hole mass $M$:
\be
H = {c}/{l_{osc}} = h_P \left(\frac{m_P}{M}\right) \quad \Lambda = {\lambda_P} \left(\frac{m_P}{M}\right)^2, \quad h_P = c /l_P, \quad \lambda_P = 3 h_P^2 /c^4
\ee
$L_{QG} = (L_G + L_Q)$ in Eq.(\ref{dS}) is the complete length allowing to describe both the classical, semiclassical and quantum (trans-Planckian) gravity domains. The complete  vacuum density $\rho_{QG}$ in the quantum gravity regime including the classical and quantum ones ($\rho_G$, $\rho_Q$, ), ($\rho_P$ 
being the Planck density scale), is:
\begin{equation}
\rho_{QG}= \frac{\rho_G}{[\; 1 + \rho_G / \rho_P\; ]^2} 
= \frac{\rho_Q}{[\; 1 + \rho_Q / \rho_P \; ]^2}, 
\end{equation}
$$
\rho_{QG}\; (\rho_G)  =  \rho_{QG}\; (\rho_Q) = \rho_{QG} \;(\rho_P^2 / \rho_G)
$$ 
\begin{equation}\label{rhoH}
\rho_G = \rho_P \left(H / h_P\right)^2 = \rho_P \left(\Lambda/\lambda_P \right),\;\;  \rho_P = {3 \; h_P^2}/{8 \pi G}
\end{equation} 
\begin{equation}\label{rhoQ}
\rho_{Q} = \rho_P \left({H_Q}/{h_P}\right)^2 = \rho_P  \left({\Lambda_Q}/{\lambda_P}\right) = {\rho_P^2}/{\rho_G}
\end{equation}          

The $QG$ magnitudes are complete variables covering both classical and quantum, Planckian and trans-Planckian, domains.  
 The high density $\rho_Q$ and $\Lambda_Q$ describe the quantum trans-Planckian vacuum. This is precisely expressed by Eqs.(\ref{ULambda1})-(\ref{UQ1}) applied to this case: 
\begin{equation} \label{LambdaHvalue1} 
\frac{\rho_G}{\rho_P}  = \left (\frac{l_P}{L_{G}}\right)^2 = \left(\frac{m_P}{M}\right)^2 = \left(\frac{S_Q}{s_P}\right)
\end{equation}
\begin{equation} \label{LambdaQvalue2} 
\frac{\rho_Q}{\rho_P} =  \left(\frac{l_P}{\Lambda}\right) = 
\left(\frac{M}{m_P}\right)^2 = \left(\frac{S_G}{s_P}\right) 
\end{equation}
The last r.h.s. of Eqs.(\ref{LambdaHvalue1})-(\ref{LambdaQvalue2}) show the link to the 
gravitational {\it entropy}: quantum 
 gravitational $S_Q$ and classical/semiclassical $S_G$ entropy. 

 ($\Lambda$, $\rho_G$)   describe a {\it classical gravitational vacuum}: a empty or dilute gravitational vacuum state of {\it large classical} sizes $L_G = l_P\sqrt{\lambda_P/\Lambda} = l_P~(M / m_P) $, very small density and  very low $\Lambda$ values.  Consistently, the {\it small} value of the quantum gravitational entropy $S_Q$ is equal to such small $\Lambda $ value.

($\Lambda_Q$, $\rho_Q$) describe a {\it quantum gravitational vacuum}, truly in the trans-Planckian domaine of very small sub-Planckian sizes $L_Q = l_P \sqrt{\Lambda /\lambda_P} = l_P~ (m_P / M)$, very high density and very high $\Lambda_Q$ values. Consistently, the  {\it high} value of the classical gravitational entropy $S_G$ is equal (in Planck units) to such high $\Lambda_Q$ value. 

\bigskip

The external black hole region is precisely a {\it classical gravity   dilute vacuum}, which in the present universe cannot be larger than the observed very low values of the classical cosmic vacuum density and cosmic vacuum energy $(\Lambda,\rho_G)$  \cite{Riess},\cite{Perlmutter},\cite{Schmidt},\cite{DES},\cite{Planck6}. The quantum duals of the classical present universe cosmic vacuum values provide an upper bound to the high values $(\Lambda_Q,   \rho_Q)$ in the quantum central vacuum black hole region as determined by Eqs.
(\ref{rhoH})-(\ref{LambdaQvalue2})

\medskip

We quantize the $(X, T)$ dimensions  which are relevant 
to the quantum space-time structure.
The remaining spatial transverse dimensions $ X_{\bot} $ are considered here 
as non-commuting coordinates. 
This corresponds to quantize the two-dimensional surface $(X,T)$ relevant 
for the light-cone structure. Notice that although the transverse spatial dimensions ${\bot}$ have zero commutators
they could fluctuate. This is enough for considering the 
novel features arising in the  quantum space-time structure and the {\it quantum light cone}.

\section{Concluding Remarks}

\bigskip

(i) This approach is a first step to cover globally and non-perturbatively the classical, semi-classical and quantum gravity domaines of black holes. This  framework supports and is consistent with the idea that a quantum theory must {\it be finite}. The global $QG$ variables and quantum discrete space-time  include here the  highly quantum 
trans-Planckian domaine and go  well beyond  other approachs.

\medskip

(ii) The trans-Planckian domaine in black holes  {\it is found in the central interior region}, and this is so for {\it all} black hole masses, including astrophysical and macroscopical black holes which exterior space-times are  classical and semiclassical regions. The highly excited vacuum central region is a constant curvature de Sitter vacuum without any singularity. The most central quantum trans-Planckian black hole region have the most higher excited levels, with $ \Lambda_{Qn} = M_n = \sqrt{ 2n+1 }$ (in Planck units)  and 
 smallest quantum gravitational entropies $S_{Qn} = 1/ (2n + 1)$. 

\medskip

(iii) De-excitation of the levels go from the central quantum  trans-Planckian core of the black hole with high $n$  until $n = 0$ (the Planck scale), and then entering the semiclassical/classical gravity exterior space-time region, more and more de-excited and classical for increasing $n$, (the classical branch), with decreasing vacuum energy and a continuum spectrum
reaching asymptotically  flat space-time.  
In the process of classicalization, $n$ increases from the Planck level $(n = 0)$, $X_n = \sqrt{(2n+1)}$ increases,  the huge and finite values of the central black hole vacuum energy and curvature diminish as $ 1/(2n+1)$, and vanish asymptotically for very huge $n$. 
This is coherently accompassed by the increasing distances $L_n = \sqrt{(2n+1)}$, and the
increasing  levels $S_{G n} = (2n+1)$ of the 
 Bekenstein-Hawking entropy which is a classical/semiclassical gravitational entropy,  
and it is always un {\it upper bound} to the other entropies. 

\medskip

(iv) Recall that quantum back reaction 
effects, gravitational scattering near a event 
horizon structure produces a quantum shift too (the shifted horizon)  \cite{tHooft}, \cite{dVNS1988}, 
\cite{Sanchez1987}. This approach consistently describe too the cosmological phases from the pre-Planckian or trans-Planckian quantum phase to the Planck scale and then to the post-Planckian universe: Refs \cite{NSPRD2021}, \cite{Sanchez3}.

\medskip

(v) The identification of space-time ("IST") have been investigated in the past and recent years at the level of semiclassical gravity \cite{SanchezCargese1987}, \cite{NGSWhitingNPB1987}, \cite{NGSDomechLevinasIJMPA1987}, \cite{tHooft1}, 
 \cite{StraussWhitingFranzenCQG2020}, \cite{tHooft2022}. In our framework here we have not used  IST, but as already pointed in \cite{NSPRD2021}, \cite{Sanchez2019}, 
our results support  CPT  and IST in the full quantum theory. In semiclassical gravity, the  symmetric (or antisymmetric) IST QFT provide a CPT symmetry of the theory.  In the euclidean (imaginary time) manifold, the differences between the four Kruskal space-time regions dissapear and they became automatically identified. And in the central trans-Planckian region of the hyperbolic (real time) quantum space-time, the four Kruskal regions merge into one single region and became automatically identified.

Other approachs to the black hole interiors have been considered recently, see for example \cite{tHooft2022}, \cite{frolov},  \cite{ottewill}. In Refs \cite{brustein1},  \cite{brustein2} a regular black hole interior is described classically with a classical space-
time geometry sourced by a maximal negative radial pressure. Interestingly, (e.g in \cite {frolov} and refs therein), the black  
hole interior model is regular too with a de Sitter like geometry.  These  are effective 
like models and could help too to study to disentangle the properties  of the black hole 
interiors through different observational gravitational signals.

\medskip

In our work here, the black hole interior does appear as a fully quantum gravity region. Interestingly enough, this feature  also appears from a different approach  using scaling arguments in maximal entropic  states eg Ref \cite {brustein3} ,  which shows the consistency of the results. 
In our paper here, such feature is a direct consequence of the classical - quantum gravity duality, which provides in addition that the black hole interior is necessarily trans-Planckian.  And from a fully quantum space-time description (a quantum algebra of non-conmutative space-time instead of a space-time metric) we find that the interior is totally regular and of constant curvature. This provides the picture that the black hole interior is a truly quantum trans-Planckian vacuum, totally regular and of constant curvature. In addition, the quantum Penrose diagram is new and had not been considered before, as well as the quantum completion of the Gibbons-Hawking instanton, with the quantum trans-Planckian core at the black hole center .
These results allow better describe and understand  the total regularity of the quantum black hole space-time, eg the non-singularity at the center, the description of such interior and exterior regions and their connection to the constant curvature vacuum  describing dark energy.  
The complete partition function is new and allows to understand the discrete spectrum of the different black hole regions, accompassed by the complete entropy and black hole evaporation stages. 

Is not our aim here  to discuss a review of the black hole interior literature. Our work here is in the context of trans-Planckian physics that does appear necessary to describe the black hole interiors, which classical gravitational dual provide the black hole exteriors, and thus a global unifying description of the space-time is provided, the same approach allows the description of the very early cosmological phase before inflation, with its classical gravitational dual (today dark energy).

\section{Conclusions}

\begin{itemize}

\item{ Overall, a consistent  
 quantum picture of the black hole  space-time does appear from  the internal central black hole regions which are the most quantum and trans-Planckian, to the semiclassical and classical external regions 
until the asympotically flat far regions from the black hole, together with  their 
physical magnitudes and spectrum: size, mass, partition function,  gravitational
entropies and temperatures covering all mass range and gravity domains:  quantum  (trans-Planckian) gravity  and semiclassical/classical gravity domains.}   

\item{ The quantum vacuum energy bends
the space-time and produces a constant curvature background in the central black hole region of  Planck length  radius $l_P$.  We find the quantum discrete
levels: length, mass vacuum energy,
and gravitational entropy and temperature from the black hole 
 central trans-Planckian vacuum, passing through the Planck scale, to the external semiclassical and  classical exterior vacuum regions.  The gravitational entropy of the Universe today $S_{today} = (2n+1) = 10^{122}$\;is the  absolute upper bound to all entropies, in particular to all black hole entropies.}

\item{The quantum space-time structure 
allows a
{\it new quantum region} which is purely quantum
vacuum or zero-point Planckian and trans-Planckian energy and constant curvature. This central quantum vacuum core is 
a de Sitter quantum trans-Planckian vacuum  described through the relevant quantum non-commutative coordinates and the quantum hyperbolic structure.}

\item {In the external black hole space-time, the  discrete levels
 extend from the Planck scale level $(n = 0)$ and low $n$ to the quasi-classical and classical levels (intermediate and large $n$), tending asymptotically  (very large $n$) to a classical continuum space-time. Consistently, these levels have larger gravitational 
 (Gibbons-Hawking) entropy $S_{Gn} = (2n + 1),\; n = 0, 1, 2, ...$ and lower vacuum energy $\Lambda_{n} =  1/(2n + 1)$. In the central quantum trans-Planckian core of the black hole, 
  the levels extend 
 from the Planck scale $(n = 0)$ to the lengths smaller than the Planck scale, until the quantum 
highly excited trans-Planckian levels (very large $n$) which are those of smaller entropy $S_{Qn} = 1 / (2n + 1)$ and higher vacuum density $\Lambda_{Qn} = (2n + 1)$.}

\item{There is {\bf no singularity} at the black hole origin. First: the  $r = 0$ mathematical singularity is
{\bf not} physical: it is the result of extrapolation of the purely
classical (non quantum) General Relativity theory, {\it out of 
its domain of physical validity}. The Planck scale is not merely 
a 
useful system of units but a physically meaningful scale: the 
onset of quantum gravity; this scale precludes the extrapolation 
until 
zero time or length. This is precisely what is expected from 
quantum trans-Planckian physics in gravity: the smoothness of the 
classical gravitational 
singularities. Second: de Sitter vacuum which is the vacuum 
interior region of the black hole is a smooth constant curvature 
vacuum {\bf without any curvature singularity}. Third: the small 
and a trans-Planckian  vacuum have a  high but {\it bounded} 
trans-Planckian constant curvature and therefore
 {\it without singularity}.}

\item{There are {\it no} singularity boundaries at the quantum level at $(T^2 - X^2)(r = 0) = \pm 1 $  
nor at $(T^2 - X^2) = \pm \sqrt {2}$ . 
The quantum space-time {\it extends} without boundary beyond the Planck hyperbolae
$(T^2 - X^2)(n=0)  = \pm 1 $ towards {\it all} levels. $(T^2 - X^2) = \pm \sqrt {2}$ are the quantum hyperbolae which 
replace the classical singularity:  $(T^2 - X^2)_{classical} (r=0) = \pm 1$. Moreover,
the quantum hyperbolae $(T^2 - X^2) = \pm \sqrt {2}$  lie {\it outside} the 
allowed quantum hyperbolic levels $(T^2 - X^2)_{n} = (2n + 1)$, $n = 0, 1, 2, ...$, and therefore they {\it are excluded} at the quantum level: The
singularity is {\it removed} out from the quantum space-time. There is {\it no}  singularity boundary in the quantum space-time, not at $r = 0 = x*$, not at any other place. The quantum Schwarzschild - Kruskal space-time is 
 {\bf totally regular}. }

\item{The quantum trans-Planckian core is present in {\it all} black holes,  macroscopic and astrophysical ones. In the imaginary time manifold (instanton), it appears too, and allows to complete the classical gravity Gibbons-Hawking instanton, which is cutted at the horizon: The classical black hole instanton is thus regular but {\it not} complete. The black hole quantum instanton is regular and complete. The  complete partition function, temperature and  entropy all reflect this feature and clearly include the highly excited and dense trans-Planckian central region of radius $l_P$, as well as the discrete levels, density of black hole states and black hole decay rate.}

\item{States with the Planck mass $m_P$ are {\it not} black holes, they are enterely quantum gravity states, decaying in 
the way heavy particles or quantum strings do, in this case in gravitons, other elementary particles and radiation. Black holes reaching the Planck mass in the process of their evaporation undergo a phase transition into a pure (non mixed) quantum state which decay in gravitons, particles and radiation.}     

\item{The results of this paper could provide insights 
for research directions and new 
understanding in quantum theory and gravity and for the searching of
quantum gravitational signals, for e-LISA \cite{LISA} for 
instance, after the 
success of LIGO \cite{LIGO},\cite{DESLIGO},
as well as for other quantum signals in space-
time, \cite{Simons}, \cite{Euclid}, \cite{DESI}, \cite{WFIRST}, black holes in particular, for astrophysical black holes and for 
"quantum black holes", or the last stages and "remnants" of black hole evaporation and  black hole "explosions". One of the  
 novel results of this paper is that quantum 
 physics is a inherent constituent of  {\bf all} black 
 hole interiors, from the horizon to the center, in particular in the most larger and astrophysical black holes. It is a result of this paper too that the black hole interior trans-Planckian vacuum is of the same nature of the very early cosmological vacuum: quantum, trans-Planckian and of constant curvature, which classical gravity dual is a very dilute, very low energy gravitational vaccum (today dark energy).}
\end{itemize}

\newpage

{\bf ACKNOWLEDGEMENTS}

\bigskip

The author acknowledges 
discussions and communications with Gerard 't Hooft, Adam Riess 
and Roger Penrose on various occasions.

\end{document}